\documentstyle[floats,psfig,aps]{revtex}
\begin{document}
\twocolumn
\title{Charmonium suppression in $p\,A$ collisions}

\author{F. Arleo, P.-B. Gossiaux, T. Gousset and J. Aichelin}

\address{SUBATECH \\
Laboratoire de Physique Subatomique et des Technologies Associ\'ees\\
UMR Universit\'e de Nantes, IN2P3/CNRS, Ecole des Mines de Nantes\\
4, rue Alfred Kastler,
F-44070 Nantes Cedex 03, France.}

\date{\today}
\author{\begin{quote}
\begin{abstract}
The new high precision data on charmonium production in proton-nucleus
collisions by the E866/NuSea collaboration at Fermilab allow ---
together with older data at lower energies --- to fix a unique set of
parameters for the standard production and absorption scenario of
charmonium in a proton-nucleus reaction. In this scenario the
$c\bar{c}$ pair is formed in an octet state, emits a gluon and
continues its radial expansion in a singlet state until it has reached
the charmonium radius. In all three phases it can interact with the
nuclear environment. We find that the lifetime of the octet state is
much shorter than acceptable on physical grounds. This challenges the
physical reality of the first phase in the standard scenario.
\end{abstract}
\end{quote}}
\maketitle

\section{Introduction}
Since the suppression of $J/\psi$'s has been proposed by Matsui and
Satz as a signal for a quark-gluon plasma formation more than a decade
ago~\cite{mat86} a lot of experimental efforts have been devoted to
use them for the quest of the formation of a quark gluon plasma
created in ultra-relativistic heavy ion collisions. For small systems,
where the creation of a plasma is not expected, the measured
absorption can be well explained without invoking a plasma.  At Quark
Matter 1996, the first results on the heavy system Pb-Pb have been
reported~\cite{gon96} by the NA50 collaboration and it has been argued
that they show an abnormal suppression (as compared to the
extrapolation from small systems) which may be interpreted as a hint
for the formation of a quark gluon plasma. This triggered a lively and
still ongoing debate about the conclusions of this observation.

Before one tries to understand the $J/\psi$ suppression seen in heavy
ion collisions it is necessary to understand that obtained in $p\,A$
collisions where the reaction is much more under control. There one
expects that the shape and the density of the nucleus does not change
during the reaction. Recently new data have been advanced by the
Fermilab group E866/NuSea~\cite{lei98}, where for the first time
differences between the absorption of $J/\psi$'s and $\psi'$'s have
been observed. This fact as well as the high precision of these data
allow to fix the parameters of the standard
scenario~\cite{bla89,vog91}.

Here, we present a model for the charmonium absorption in
proton-nucleus collisions which is based on the standard scenario
described in Section~II. We have limited the ingredients to the
minimum that allows the description of the Fermilab data in the whole
$x$ range. 
Several initial state effects such as gluon shadowing \cite{arn94},
energy loss \cite{bh,gm}, or quantum coherence \cite{kha93}
could affect somehow charmonia production at Fermilab energies.
However, since Drell-Yan production does not show a significant nuclear
dependence, we have assumed that those effects 
are relatively less important and their consideration would
imply an unnecessary complication. We have checked that inclusion of
gluon shadowing does not alter the conclusions one can draw from our
study, but changes somewhat the present set of effective
parameters. Taking the recent results of the E866/NuSea collaboration
at Fermilab~\cite{lei98}, we determine a set of parameters for the
standard scenario. The same set of parameters allows the description
of almost all other available data on charmonium suppression in $p\,A$
collisions, including the majority of NA3 data points~\cite{bad83}, those of
NA38~\cite{abr99} and those of E772~\cite{ald91}. We show that the
functional form of the model allows to describe the whole body of data
and that its parameters are rather precisely fixed by them. We show as
well that the $x_2$ scaling predicted by the model is compatible with
high-energy data. We do not claim that our result excludes other
reaction mechanisms (as we will see there is good evidence for that on
physical grounds) but stress that on the basis of the presently
available data one cannot distinguish experimentally between them.

\section{Model for Charmonium Absorption in proton-nucleus reactions}

In $p\,A$ collisions the production of $c\bar{c}$ states is supposed
to be proportional to the atomic mass $A$ of the target, as it is the
case for the Drell-Yan process. Thus, the charmonium states are
produced anywhere in the nucleus with a probability directly
proportional to the local nucleon density. We consider here
$J/\psi$'s, $\psi'$'s and $\chi$'s. Feeding is important: only 50\% of
the observed $J/\psi$'s are primarily $J/\psi$'s, the other 50\% come
from radiative decays of $\chi$'s (43\%) and $\psi'$'s
(7\%)~\cite{kha93}. The spin of the $\chi$ state may play a role in
the $\chi\,N$ cross section~\cite{ger98}, but we neglect this effect
here.

The $c\bar{c}$ pair is not created in a fully formed state, with its
final radius. We assume that the transverse distance between the
quarks increases linearly with time. In the $c\bar{c}$ rest frame we write
\begin{equation}\label{radius}
r_{c\bar{c}}(\tau) = 
\left\{
\begin{array}{rl}
r_{0}+v_{c\bar{c}}\; \tau  &
\qquad {\mathrm{if}}\ r_{c\bar{c}}(\tau) \leq r_{i},\\
r_{i}	\qquad	 & \qquad {\mathrm{otherwise,}}\\
\end{array}
\right.
\end{equation}
where $i$ stands for $J/\psi$, $\psi'$ and $\chi$. The final radii
$r_{i}$ of the different states are determined by microscopic
calculations. The limits chosen are $0.43$~fm for $J/\psi$, $0.87$~fm
for $\psi'$ and $0.67$~fm for $\chi$~\cite{gos93}. $v_{c\bar{c}}$ is the sum of 
the $c$ and $\bar{c}$ velocities i.e. $v_{c\bar{c}} = 2 v_{c}$.
$v_{c\bar{c}}$ and the initial
radius $r_{0}$ are taken as free parameters. An alternative
possibility for the $c\bar{c}$ expansion, $r\propto\sqrt{\tau}$, has
been proposed in~\cite{far88}. We have checked that no definite
conclusion may be drawn about which one is actually at work. The main
point is that the expansion is important in the kinematical region
where one can distinguish $J/\psi$ from $\psi'$.

The $c\bar{c}$ pair is formed by two-gluon fusion in an octet state
$(c\bar{c})_{8}$ which later on emits a gluon and neutralizes its
color. We assume that the emission occurs after a color neutralization
time $\tau_{8\to 1}$ in the rest frame of the $c\bar{c}$ pair,
independently of its velocity. $\tau_{8\to 1}$ is the third parameter
of the model. In the nucleus rest frame, the lifetime of the octet
state increases with $x_{F}$, due to Lorentz dilatation. We will see
that this fact is responsible for the increase of absorption with
$x_F$, as seen in data. This would not be the case if the
octet-singlet transition occurred after a constant time in the nucleus
rest frame (a natural hypothesis if neutralization was caused by
interactions with the surrounding nucleons). According
to~\cite{kha93,kha94} the octet lifetime should be of the order of
$0.3$~fm. We will see that the new data allow for a precise
determination of the color neutralization time $\tau_{8\to 1}$.

Some authors have argued that the production can partly proceed via a
direct color singlet state~\cite{won96}. Although a priori distinct
from color evaporation they have shown that it is not presently
possible to clearly disentangle those two possibilities. Since on the
one hand the $x_{F}$ dependence of the color singlet production
fraction is unclear and on the other hand the fit does not improve if
one incorporates this effect we assume that $c\bar{c}$ pairs are
always created in a color octet state.

The singlet-nucleon absorption cross section $\sigma_{(c\bar{c})_1N}$
depends on both the transverse radius $r_{c\bar{c}}$ of the $c\bar{c}$
pair and its energy. In QCD the total cross section for a compact
singlet state should be proportional to the square of its
radius~\cite{low75} and we write $\sigma_{(c\bar{c})_1N}=\sigma_{\psi
N}(s) \cdot(r_{c\bar{c}}/r_\psi)^2$. Next, we need the energy
dependence of the $\psi N$ cross section in the kinematical region
corresponding to available data, that is in the region around 10~GeV
and above. In this range it may be related to that for $J/\psi$
photo-production and to the small $x$ gluon distribution~\cite{huf98}
leading to $\sigma_{\psi
N}(s)=\sigma_1\cdot(\sqrt{s}/10{\mathrm{~GeV}})^{0.4}$ where $\sqrt s$
is the center of mass energy of the $c\bar{c}N$ system. Thus, the
time dependent cross sections for singlet states is given by
\begin{equation}
\sigma_{(c\bar{c})_1N}(\tau) =\sigma_{1}\cdot 
\left(\frac{\sqrt{s}}{10{\mathrm{~GeV}}}\right)^{0.4} 
\left(\frac{r_{c\bar{c}}(\tau)}{r_{\psi}}\right)^{2},
\label{cross1}
\end{equation}
where the radius $r_{c\bar{c}}$ is given by (\ref{radius}) and
$\sigma_{1}$ is the cross section for a fully formed $J/\psi$ with
incident energy $\sqrt{s}=10$~GeV.  $\sigma_{1}$ is the fourth
parameter of our model and should be of the order of a few
mb~\cite{huf98}.

The present model gives a unified treatment for $J/\psi$, $\psi'$ and
$\chi$ since the only difference in the cross sections comes from the
final radii $r_{i}$. Consequently, one expects for small $x_F$ where
the $J/\psi$ is formed before leaving the nucleus a difference in the
absorption between $J/\psi$, $\psi'$ and $\chi$, as the $\psi'$ and
the $\chi$ radii are bigger than that of the $J/\psi$. Indeed the
E866/NuSea group found a different absorption for $x_{F}\le 0.2$
(cf. Figs.~\ref{fewbe} and~\ref{alpha}). At high $x_{F}$, the
different states will behave similarly because the $c\bar{c}$ leaves
the nucleus before it has reached the $J/\psi$ final size, due to
Lorentz dilatation. Strictly speaking this is true if one neglects the
mass difference between charmonia. For given beam energy and
charmonium velocity $x_F$ is proportional to the charmonium mass
implying, e.g., a systematic 20\% shift of $\psi'$ with respect to
$J/\psi$. This is somewhat reduced by the fact that $J/\psi$ comes
partly from radiative decay of $\chi$ and $\psi'$ and further by the
fact that what one needs in order to compute the $c\bar{c}$ velocity
at a given $x_F$ is the intermediate $c\bar{c}$ mass. In lack of a
good description of this mass effect and noticing that present $\psi'$
data are not precise enough to be analyzed at this level of accuracy,
we assumed that the intermediate $c\bar c$ and all charmonia have the
same mass $m_{c\bar{c}}=(m_\psi+m_{\psi'})/2=3.4$~GeV.

The interaction between the octet state and nucleons is mainly
responsible for the increase of absorption with increasing $x_{F}$
observed at Fermilab. The large $x_{F}$ region is dominated by $c\bar
c$ pairs which have left the nucleus in the octet state. The
absorption cross section between octet states and nucleons
$\sigma_{(c\bar{c})_8N}$ may be of the order of 20 mb \cite{kha93}, that is much
larger than the singlet cross section. The usually quoted value of 6 mb in
hadroproduction would be interpreted in the present scenario as 
an average between singlet and octet cross sections.
Since the energy dependence is related to gluon dynamics, we assume
that it is the same for 
$\sigma_{(c\bar{c})_8N}$ and $\sigma_{(c\bar{c})_1N}$. The octet
cross section is taken to be independent of the radius of the pair
$(c\bar{c})_{8}$:
\begin{equation}
\sigma_{(c\bar{c})_8N}=\sigma_{8}\cdot
\left(\frac{\sqrt{s}}{10{\mathrm{~GeV}}}\right)^{0.4}.
\end{equation}
$\sigma_{8}$ is the fifth and last parameter of the model. 

If one assumes singlet states only, the $A$ to $p$ production ratio
approaches 1 at $x_{F}=1$ because the large gamma factor between the
$c\bar{c}$ pair and the laboratory frame does not allow an expansion
of the $c\bar{c}$ system while traveling through the nucleus. Thus the
cross section (\ref{cross1}) is small and so is the absorption.

\section{Results}

Quantitative results were obtained with a Monte-Carlo simulation. The
computation is done by first generating $3A$ random numbers, the first
$3(A-1)$ describing the positions of $A-1$ static nucleons inside the
nucleus and the 3 last localizing the creation point of the charmonium
inside the nucleus. From this creation at the proper time $\tau=0$, we
follow the charmonium on its way through the nucleus in the $z$
direction, $z(\tau)=z_0+\beta\gamma\tau$. The charmonium velocity
$\beta$ in the nucleus rest frame is determined by $x_F$ (see
Section~IV). The charmonium neutralizes its color at time $\tau_{8\to
1}$ and expands as expressed by (\ref{radius}). If the transverse
distance between the charmonium and a nucleon becomes smaller than
$\sqrt{{\sigma(\tau)/\pi}}$ we assume that the charmonium is absorbed.

We made a fit to the $J/\psi$ suppression in tungsten normalized to
beryllium, measured by the E866/NuSea collaboration~\cite{lei98}, with
the five parameters described in the previous section. We obtain the
following parameters: $r_{0}=0.15$~fm, $v_{c\bar{c}}=1.85$,
$\sigma_{1}=2.1$~mb, $\tau_{8\to 1}=0.02$~fm and
$\sigma_{8}=22.3$~mb. This set of parameters is used to calculate the
results for all other energies and targets. The influence of a change
of these parameters on the results is discussed in Section~VI.

\begin{figure}[htbp]
\centerline{\psfig{figure=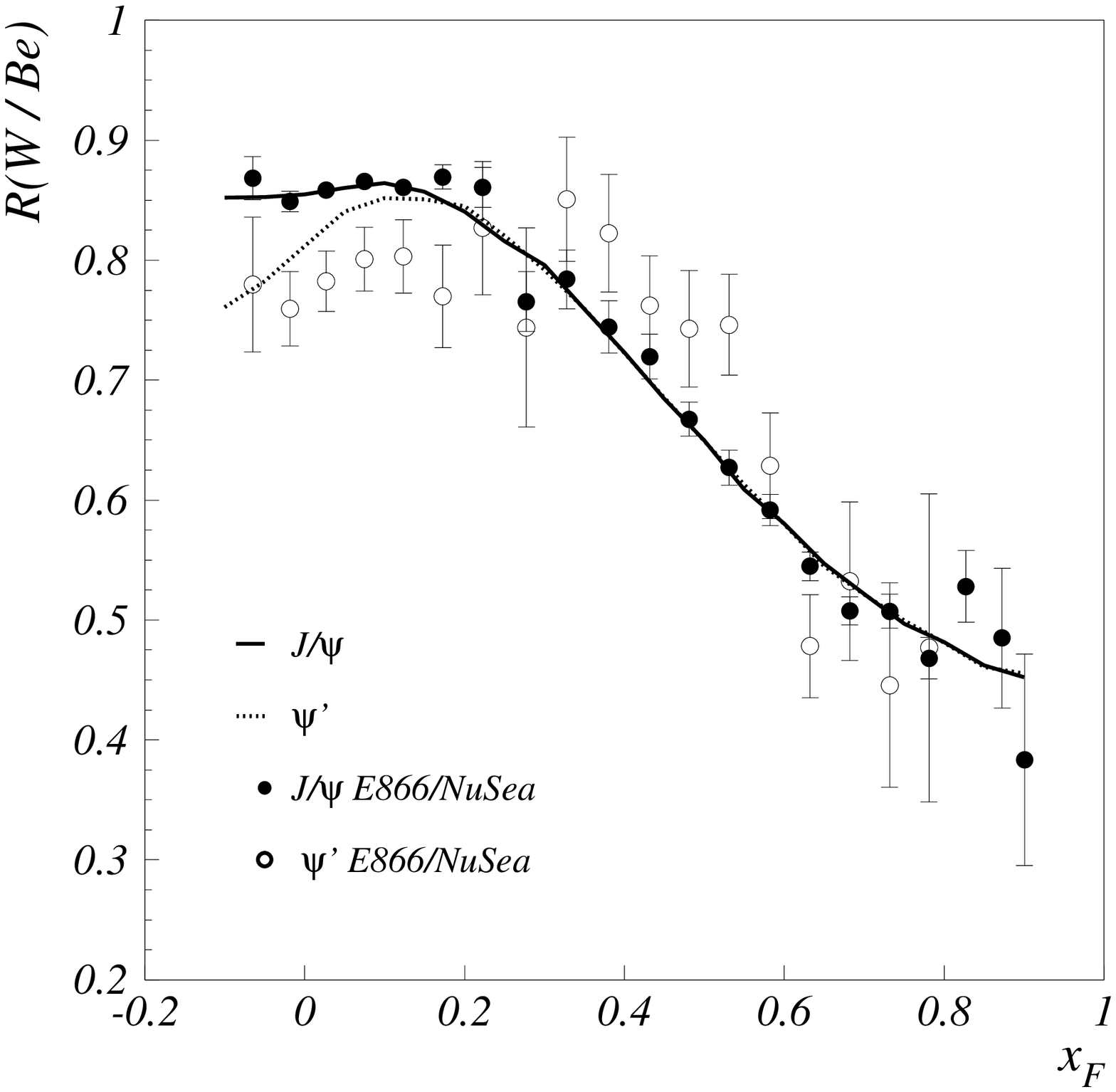,width=\hsize}}
\centerline{\psfig{figure=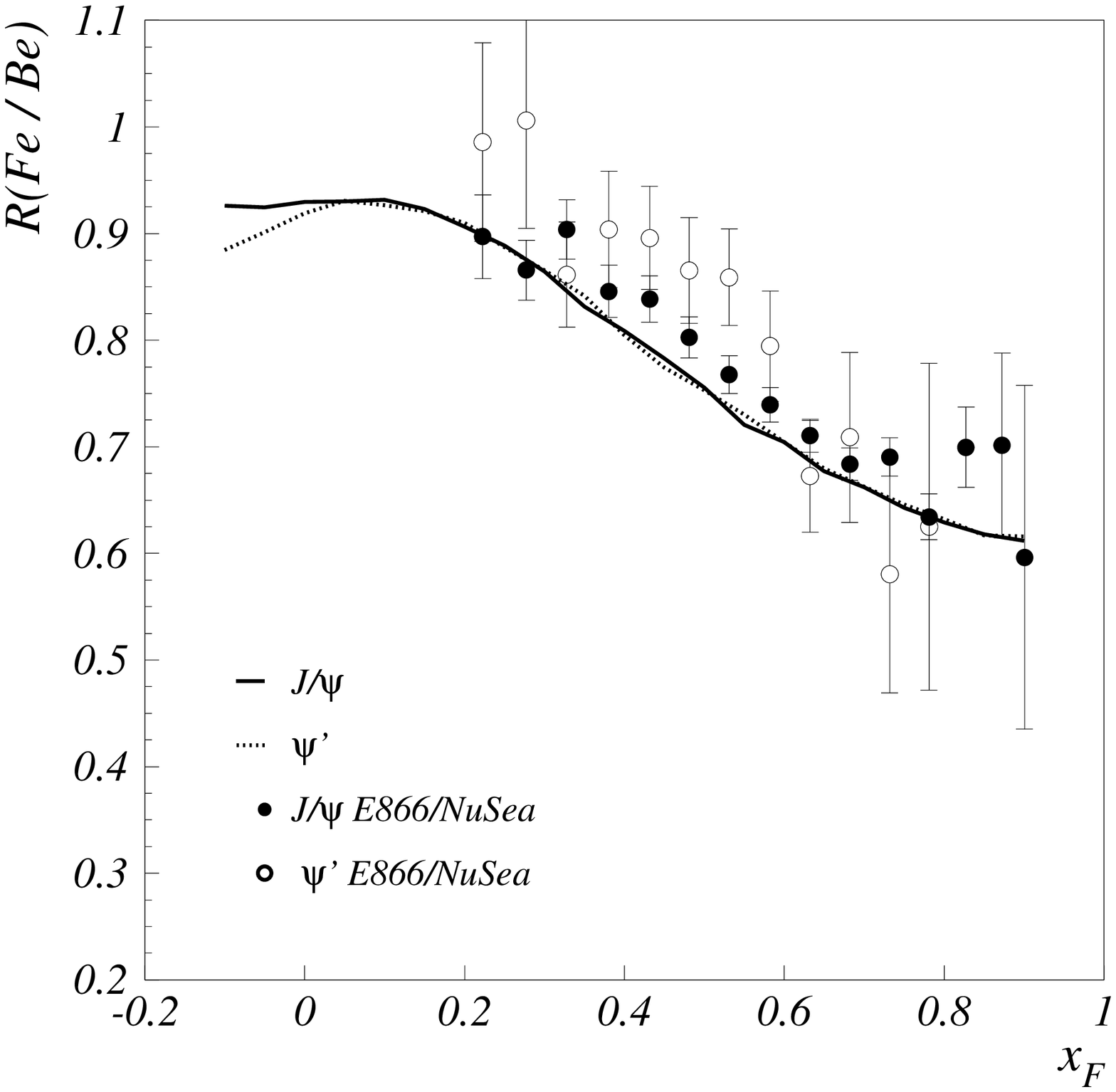,width=\hsize}}
\caption{Ratio
$R(A/{\mathrm{Be}})=(9/A)\cdot\sigma(A)/\sigma({\mathrm{Be}})$ versus
$x_F$ of produced $J/\psi$'s (black circles) and $\psi'$'s (white
circles) for W (top) and Fe (bottom). The lines are the results of our
model.}
\label{fewbe}
\end{figure}

In Fig.~\ref{fewbe} we compare our calculation with the recent
E866/NuSea results of proton-nucleus reactions at
800~GeV. In this experiment the absorption ratio
\begin{equation}
R(A/{\mathrm{Be}})=\frac{9\,\sigma(A)}{A\,\sigma({\mathrm{Be}})},
\end{equation}
for W and Fe has been measured. First of all we see that below
$x_F=0.2$ the absorption of $J/\psi$ and $\psi'$ is different. It is
this fact which allows to determine the parameters of our model
precisely. We see as well that the absorption has a minimum around
$x_F = 0.1$. As we will discuss later in detail the increase at large
$x_F$ is caused by octet state-nucleon collisions whereas that at
negative $x_F$ comes from collisions of the fully formed charmonium
states with nucleons, thus explaining the difference between $J/\psi$
and $\psi'$ absorption. The above parameter set describes data quite
well in the whole measured $x_F$ range and for the two different
ratios (notice that the fit was performed with the W to Be production
ratio only).

\begin{figure}[htbp]
\centerline{\psfig{figure=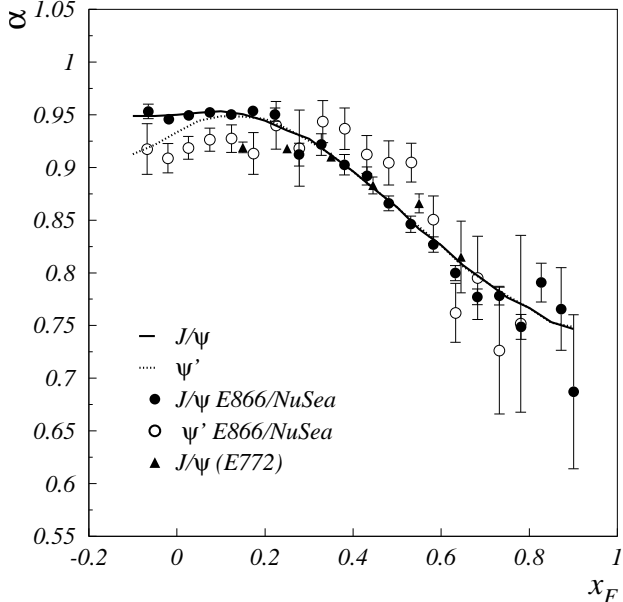,width=\hsize}}
\caption{$\alpha=1+\ln (R({\mathrm{W}}/{\mathrm{Be}}))/\ln (184/9)$ versus
$x_{F}$ from E866/NuSea for 800 GeV/c protons as compared to the
results of the calculation. E772 data are also shown for comparison.}
\label{alpha}
\end{figure}

In Fig.~\ref{alpha} we show the same result in the alternative
representation
\begin{equation}
\alpha=1+\frac{\ln R({\mathrm{W}}/{\mathrm{Be}})}{\ln 184/9}.
\end{equation}
A thorough study of this quantity~\cite{lei99} has shown that
E866/NuSea results are compatible with earlier less precise results at
800~GeV and for various nuclei~\cite{ald91}. We present them here for
completeness though in contradistinction to the E866/NuSea data they
are not corrected for $p_T$ acceptance~\cite{lei99}.

\begin{figure}[htbp]
\centerline{\psfig{figure=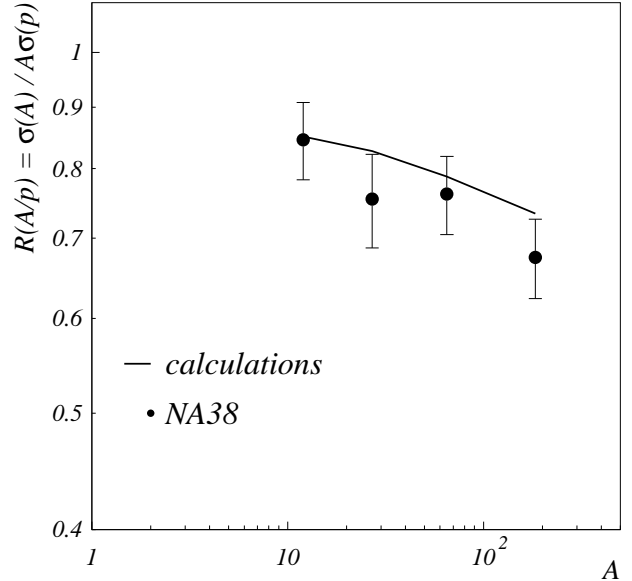,width=\hsize}}
\caption{$J/\psi$ absorption for various nuclei in $p$(450~GeV)$+A$
reactions from NA38 as compared to the calculation.}
\label{plotna38}
\end{figure}

The NA38 collaboration measured the $J/\psi$ production for several
targets (C, Al, Cu and W) at a beam energy
$E_p=450$~GeV~\cite{abr99}. The center of mass rapidity range is
$[-0.4,0.6]$ corresponding to $-0.1\le x_F\le 0.15$. The comparison
with our model is displayed in Fig.~\ref{plotna38}, with $x_{F}$ taken
as 0.05.

\begin{figure}[htbp]
\centerline{\psfig{figure=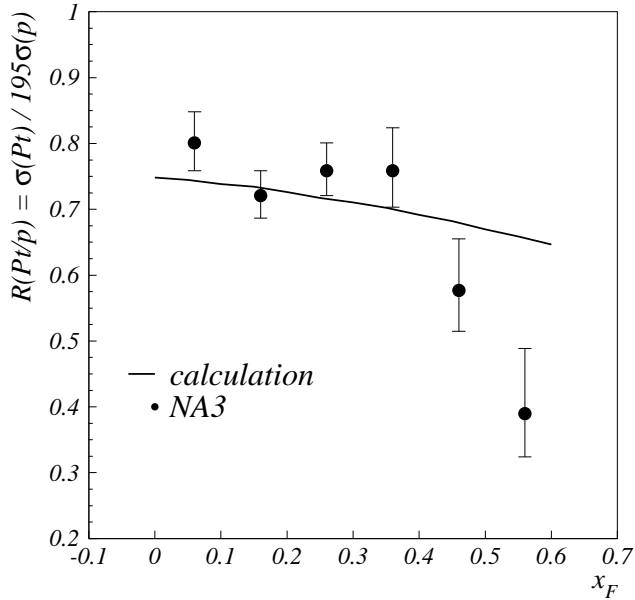,width=\hsize}}
\caption{Ratio of $J/\psi$'s produced in $p$(200~GeV)+Pt and
$p$(200~GeV)+p versus $x_{F}$ from NA3 as compared to the
calculation.}
\label{na3xf}
\end{figure}

Finally, we compare our model with NA3 data~\cite{bad83}. Measurements
were done with proton as well as pion beams on a proton and on a
platinum target, at 200~GeV. Here, one can compare data with the model
for $0\leq x_{F}\leq 0.6$. One observes in Fig.~\ref{na3xf} a fair
agreement except for the two largest $x_{F}$ values.

For the largest $x_F$ value the deviation is large and the trend shown
by large $x_F$ NA3 data points cannot be described in our model. This
may be the signal that some energy loss of the incoming gluon
manifests itself since a loss as given in Ref.~\cite{bh} would indeed
have more influence at smaller energies and larger $x_F$. We found
that the suppression seen for the large $x_F$ NA3 data points may be
explained with an energy-independent 1~GeV$/$fm energy loss of the
incoming gluons in addition to the nuclear suppression considered so
far. We have checked that such a rate gives a negligible effect at
800~GeV and that the corresponding rate for quarks is compatible with
Drell-Yan measurements at 280 GeV~\cite{cal} and 800 GeV~\cite{vas}.

There are nevertheless some weak points in this explanation:
\begin{itemize}
\item the energy dependence of energy loss is not clear~\cite{bh,gm};
\item the energy loss also affects the ratio at low $x_F$ for 200~GeV
which becomes more suppressed than seen in the data;
\item the pion beam NA3 data do not show a strong suppression of
$J/\psi$'s at large $x_F$;
\end{itemize}
which all together prevent us from drawing a firm conclusion on the
relevance of energy loss.

\emph{Is there Scaling?} Once a good reproduction of the raw data has been 
achieved, it is
natural to ask whether a more unified picture is possible, exploiting
scaling effects. In the present scenario, the absorption of charmonium
states is entirely determined by the $\gamma$ factor of the $c\bar{c}$
pair in the nucleus frame. Therefore, the model implicitly contains a
$\gamma$ scaling.

The value of $\gamma$ is directly related to the momentum fraction
$x_{2}$ in the target nucleon of the gluon which produces the
$c\bar{c}$ pair. From $\gamma=E_{c\bar{c}}/m_{c\bar{c}}$,
$E_{c\bar{c}}=x_1 E_p$ and $m_{c\bar{c}}^2=2x_1 x_2 m_p E_p$ we find
\begin{equation}
\gamma=\frac{m_{c\bar{c}}}{2 x_2 m_p}.
\end{equation} 
With $m_{c\bar{c}}$ fixed as explained in Section~II, the scaling with
$\gamma$ thus yields a scaling with $x_{2}$.

A comparison of the E772 (which are in agreement with the more precise
E866/NuSea data) with the NA3 data led to various conclusions in the
literature. In~\cite{gup92} a good evidence was found for $x_2$
scaling using pion beam data, whereas in~\cite{ald91} scaling was found in 
$x_{F}$ and not in $x_{2}$.
Since NA3 and E866/NuSea suppression ratios are consistent for $x_2 \geq$ 0.06,
all conclusions depend on the small $x_{2}$, i.e. large $x_F$, NA3 $p\,A$ data.

\section{The Significance of the different kinematical regions} 

We now discuss in more detail which of the above-mentioned processes
affect the suppression at a given $x_F$ value.

\begin{figure}[htbp]
\centerline{\psfig{figure=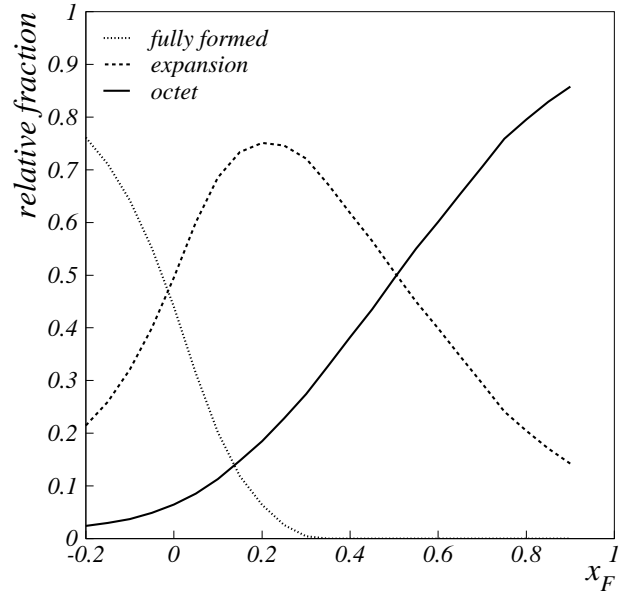,width=\hsize}}
\caption{Relative fraction of the different states, (a), (b) or (c),
of the $c\bar{c}$ pairs when leaving the nucleus in $p$(800~GeV)+W.}
\label{propsurv}
\end{figure}

Since $\tau_{8 \rightarrow 1}$ is smaller than the formation time,
one can distinguish three $c\bar c$ states: (a) a $c\bar{c}$ still in
a color octet state, (b) a $c\bar{c}$ already in the singlet state but
still expanding and finally (c) a fully formed charmonium state.
Fig.~\ref{propsurv} shows in which state the surviving $c\bar c$ pairs
leave the nucleus as a function of $x_F$ in the reaction $p$(800
GeV)+W. We see that at large $x_F$ the octet fraction is large whereas
at negative $x_F$ values almost all $c\bar c$ pairs have lost their
color at this stage.

\begin{figure}[htbp]
\centerline{\psfig{figure=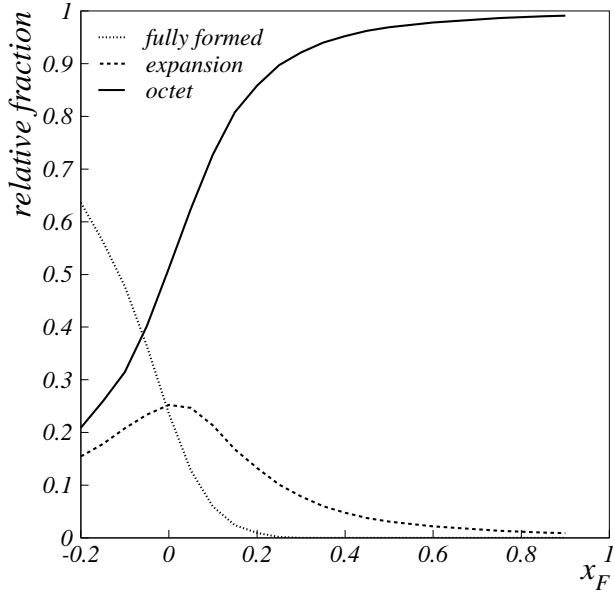,width=\hsize}}
\caption{Relative fraction of the different states of the $c\bar{c}$
pairs at the time of absorption by a nucleon in $p$(800~GeV)+W.}
\label{propabs}
\end{figure}

A complementary information is provided in Fig.~\ref{propabs}. Here we
display in which state the non surviving $c\bar c$ pairs are absorbed,
again as a function of $x_F$. This figure shows which kinematical
region is sensitive to the different cross sections and formation
times.  At large $x_F$ values the absorption rate is sensitive to
$\sigma_8$ only.  At intermediate $x_{F}$ ($x_{F}\in[0,0.2]$), one
encounters not fully formed singlet $c\bar{c}$ states. The
corresponding singlet cross section is weak, so is the associated
suppression. Therefore the absorption has a minimum in this
region. For $x_{F}<0$, the absorption is mainly governed by states in
expansion and fully formed states. Almost 50\% of $J/\psi$'s are
absorbed in a fully formed state at $x_{F}\approx -0.1$. Therefore,
differences between the charmonium states can be observed in this
kinematical region.

\section{Sensitivity to the fit parameters}

\begin{figure}[htbp]
\centerline{\psfig{figure=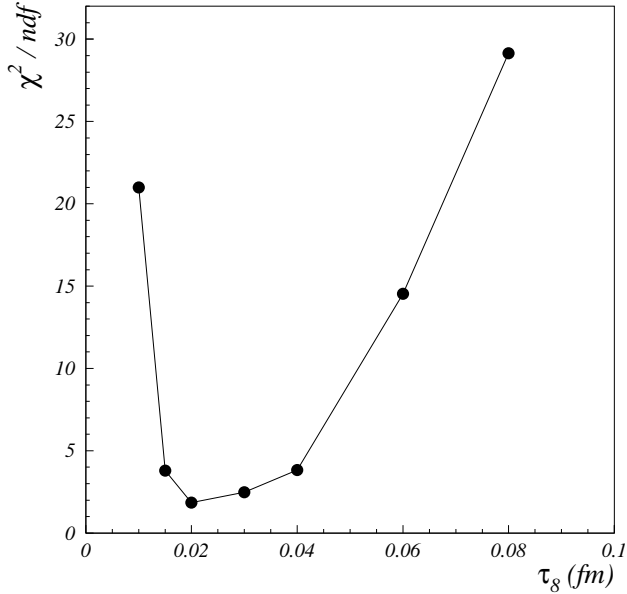,width=\hsize}}
\caption{$\chi^2/$ndf for different $\tau_{8\to 1}$. The number of
degrees of freedom is 35.}
\label{chi2}
\end{figure}

Our fit yielded a minimum at $\tau_{8\to 1}=0.02$~fm. In order to
quantify this observation we performed several fits in which
$\tau_{8\to 1}$ was fixed and the other parameters were allowed to
vary freely. In Fig.~\ref{chi2} the $\chi^2/$ndf for different values
of $\tau_{8\to 1}$ are presented. We see a rather sharp minimum around
$\tau_{8\to 1}=0.02$~fm. Thus we can conclude that the data determine
the octet lifetime precisely.

\begin{figure}[htbp]
\centerline{\psfig{figure=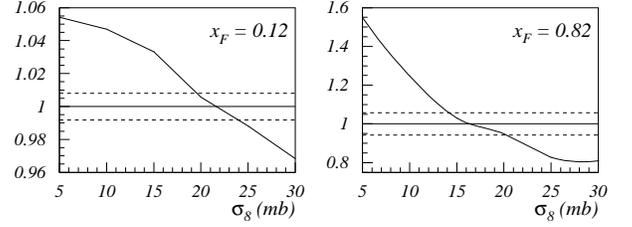,width=\hsize}}
\caption{Influence of $\sigma_8$ on charmonium absorption in W and
Be. The ratio of the calculation and the E866/NuSea data is shown for
$J/\psi$ at $x_{F}=0.12$ (left) and $x_F=0.82$ (right).}
\label{sigma8}
\end{figure}

\begin{figure}[htbp]
\centerline{\psfig{figure=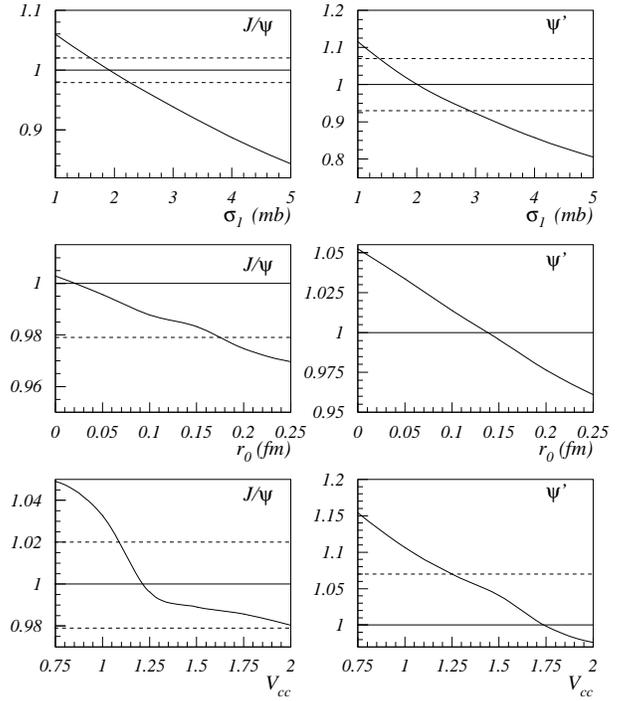,width=\hsize}}
\caption{Influence of $\sigma_1$, $r_0$ and $v_{c\bar c}$ on
charmonium absorption in W and Be. The ratio of the calculation and
the E866/NuSea data is shown for $J/\psi$ (left) and $\psi'$ (right)
at $x_{F}=-0.065$.}
\label{par}
\end{figure}

We show in Fig.~\ref{sigma8} and~\ref{par} how precisely data
determine the other parameters for $\tau_{8\to 1}=0.02$~fm. As
discussed in the former section we can almost separate two regions:
$x_F>0.4$ where the physics is determined by the properties of the
octet state and $x_F <0$ where the singlet state
dominates. Fig.~\ref{sigma8} shows the dependence of our results on
$\sigma_8$. If $\sigma_8$ decreases the absorption becomes
weaker. This is a dramatic effect for large $x_F$, which in
consequence determines this cross section quite precisely.

The variation of our results for $J/\psi$ and $\psi'$ production as a
function of the three parameters which describe the interaction of the
singlet state is shown in Fig.~\ref{par}. Here $x_F$ is fixed to
$-0.1$. We see that all these parameters are not very precisely
determined. A change of the parameters by 50\% changes the results by
about 5\%. Data at lower $x_{F}$ -for both ratios $R(W/Be)$ and $R(Fe/Be)$-
would therefore be very helpful to fix them more precisely.

The initial radius $r_{0}$ and the quark-antiquark relative velocity
$v_{c\bar{c}}$ were given by the fit as $r_{0}=0.15$~fm and
$v_{c\bar{c}}= 1.85$. Accordingly, the time needed to form a $J/\psi$
is
\begin{equation}
\tau_{f}=\frac{r_{\psi}-r_0}{v_{c\bar{c}}}\approx\; 0.15{\mathrm{~fm}}.
\end{equation}
For a longer formation time the $\psi'$'s would behave as the
$J/\psi$'s even at negative $x_{F}$ values, in contradiction to the
experimental results. A shorter formation time fails to explain the
rise of $\alpha$ between $-0.1\leq x_{F}\leq 0.1$. Calculations based
on realistic potentials for quark-antiquark bound states give a
somewhat higher value (0.3~fm) for the time of formation~\cite{kar91}.

\section{Color octet lifetime}

The fitted value $\tau_{8\to 1}$ is very small. In order to understand
the origin of such a short time in the present approach we concentrate
on the large $x_F$ Fermilab data. Here the suppression is important
and absorption in the singlet channel with a cross section of at most
a few mb plays a marginal role. Then the essential aspects are the
octet cross section and lifetime, and more precisely the comparison
between the mean free path $l_8=(\rho_0\,\sigma_{(c\bar{c})_8N})^{-1}$
and the length traversed in nuclear matter while in octet state. The
latter is bounded, on the one hand, by $\gamma\,\tau_{8\to 1}$ (the
velocity is almost 1 in the region we consider) and, on the other
hand, by twice the nuclear radius, $2R_A$. There are then two extreme
regimes:

\noindent $\;\bullet$ $\gamma\tau_{8\to 1}\ge 2R_A$. In this regime
the suppression is dictated by $R_A/l_8$ and the $x_F$ dependence
comes only from the energy dependence of $\sigma_{(c\bar{c})_8N}$ (see
Section~II). Since the $c\bar{c}N$ center of mass energy is
$2x_1E_pm_p$ and $x_1\approx x_F$ at large $x_F$, we expect a slow
dependence: $R_A/l_8\propto(x_F)^{0.2}$. This behavior is too weak to
describe the decrease seen at large $x_F$.

\noindent $\;\bullet$ $2R_A\gg\gamma\tau_{8\to 1}$. Here the
suppression is set by $\gamma\tau_{8\to 1}/l_8$. With $\gamma=x_1
E_p/m_{c\bar{c}}$, the $x_F$ dependence is now $\gamma\tau_{8\to
1}/l_8\propto(x_F)^{1.2}$. This comes close to the observed value.
The intermediate regime is an interplay of both scales.

In turn the relation between the two scales may help to fix a maximum
octet lifetime by a qualitative look at data. We see in Fig.~1 that
the $x_F$ dependence seen in data does not show any clear leveling off
except maybe in the region $x_F>0.6$ in both W to Be and Fe to Be
ratios. Taking this $x_F$ as a conservative lower limit for the
transition we deduce an upper bound for $\tau_{8\to 1}$:
\[
\tau_{8\to 1}<2R_{\mathrm{Fe}}/\gamma(x_F=0.6)=0.06{\mathrm{~fm}}.
\]

Notice that using the same reasoning with the fitted value of
$\tau_{8\to 1}$ one may estimate the value of $\gamma$ or $x_2$ at
which one sees saturation for a given $A$. For W we find
$$
x_2|_{\mathrm{transition}}=\frac{m_{c\bar c}\tau_{8\to 1}}
{2m_pR_{\mathrm{W}}}\approx 5\cdot 10^{-3}.
$$

We may also go one step further and estimate the regime at work from
the actual $x_F$ dependence seen in data. The variation of the number
of octet states in a slice of matter may be written as
$dN=-N\,dz/l_8$, i.e. $N(z)=N(z_0)\exp[-(z-z_0)/l_8]$. Combining this
indicative exponential behavior with the $x_F$ dependence explained
above for the relevant length scale and $l_8$, the octet state
suppression may be parameterized as
\[
S(x_F)=S_0^{(x_F/x_{F_0})^n},
\]
where $n$ depends on the regime at work. The value of $n$ may be
determined from data by performing the following evaluation
\[
n=x_{F_0}\frac{S'(x_{F_0})}{S_0\ln S_0},
\]
where $S'$ is the derivative of $S$ with respect to $x_F$. At
$x_{F_0}=0.6$ one finds $n\approx 1$ which ensures that the $x_F$
dependence cannot be due to the slowly varying cross section.

The value of $\tau_{8\to 1}$ necessary to reproduce data is
very short compared to several theoretical
estimations~\cite{kha95}. Such a small value has been advocated
by Wong~\cite{won96} referring to an hybrid scenario proposed by
Kharzeev and Satz~\cite{kha96} in order to cure these
conceptual problems.

The short lifetime may question, however, whether the proposed
scenario is realistic at all. The corresponding minimal energy
of the gluon emitted for color neutralization
is of the order of tens of GeV, which makes no sense. Therefore it may
very well be that the octet state is just a parameterization of a much
more complicated process. The suppression at large $x_F$ may be
attributed to a dispersion of energy in transverse direction. Such an
effect is expected if one assumes that the $J/\psi$'s are not produced
directly by gluon fusion but by fragmentation of a color string formed
between the nucleons, taking into account the string interactions
with surrounding nucleons (the probability to have these final state
interactions depending on the length of the path the string travels
inside the nucleus). This alternative is currently under a more
detailed study.

\section{Conclusion}

All presently available data including recent results from the
E866/NuSea collaboration on $J/\psi$ production in $p\,A$ collisions
can be well described in the standard color neutralization and state
expansion scenario with a common set of parameters. This is the result
of a Monte-Carlo based model in which this scenario is confronted in
detail with data.

It is questionable, however, whether this model is, as far as the
color octet state is concerned, more than a well chosen set of fit
parameters. The octet lifetime which can be precisely determined by
the new data is too short for being understandable in terms of a
physical process. Therefore it is probable that the physics observed
at large $x_F$ is more complex than assumed up to now.

This parameterization of the production and absorption of charmonium in
a hadronic environment could be extended to nucleus-nucleus
collisions. 

\acknowledgements

We are indebted to M. Leitch for exchanges concerning the Fermilab
data. We also would like to thank A.~Capella, C.~Gerschel,
M.~Gyulassy, C.~Louren\c{c}o, J.-Y.~Ollitrault, S.~Peign\'e for
valuable discussions.

\end{document}